\documentclass{DISproc}
\newcommand{\be}{\begin{equation}}
\newcommand{\ee}{\end{equation}}
\newcommand{\bea}{\begin{eqnarray}}
\newcommand{\eea}{\end{eqnarray}}

\newcommand{\bfk}{\mbox{\boldmath $k$}}
\newcommand{\bfq}{\mbox{\boldmath $q$}}

\newcommand{\pup}{p^\uparrow}

\newcommand{\kt}{k_\perp}  
\newcommand{\bkt}{\mbox{\boldmath$k_\perp$}}
\def\lsim{\mathrel{\rlap{\lower4pt\hbox{\hskip1pt$\sim$}}\raise1pt\hbox{$<$}}}
\def\gsim{\mathrel{\rlap{\lower4pt\hbox{\hskip1pt$\sim$}}\raise1pt\hbox{$>$}}}

\begin{document}
\title{Single Spin Asymmetry in $e+p^\uparrow \rightarrow e+J/\psi+X$}

\author{{\slshape R. Godbole $^1$, A. Misra$^2$, A. Mukherjee$^3$, V.
Rawoot$^2$}\\[1ex]
$^1$ Centre for High Energy Physics, Indian Institute of Science, 
Bangalore, India\\
$^2$ Department of Physics, University of Mumbai, Mumbai, India\\
$^3$ Department of Physics, Indian Institute of Technology Bombay, 
Mumbai, India} 

\contribID{xy}

\doi  

\maketitle

\begin{abstract}
We report on  a recent investigation of the single spin asymmetry (SSA) in 
low virtuality electroproduction of $J/\psi$ in color evaporation model. We
show that this can be used as a probe for the still unknown gluon Sivers 
function. 
\end{abstract}

\section{Introduction}
Large single spin asymmetries (SSA) observed when an unpolarized 
beam of electrons or protons is scattered off a transversely polarized target 
can be explained  with  the inclusion of ${\bkt}$ dependence in parton distribution 
functions (pdf's) and fragmentation functions(ff's) \cite{tmd-fact1}. 
One is led  to a generalized factorization formula called
TMD factorization \cite{Sivers1990, tmd-fact2}, which  in some
processes has been proved at leading twist and leading order \cite{fact} and has 
been argued to hold at all orders.  
The inclusion of the effect of transverse momentum of partons in parton
distribution (pdf) and
fragmentation functions leads to a new class of parton distributions
that include the effects of spin and transverse momentum
of the partons. One of these functions is the  Sivers function which
describes the probability of finding an
unpolarized parton inside a transversely polarized hadron. 
In this work, we propose charmonium production as a probe to investigate the
Sivers function and as a first step, estimate SSA in photoproduction (low
virtuality electroproduction) of 
charmonium in scattering of  electrons off transversely polarized protons.
At leading order (LO), this  receives contribution only from a 
single partonic subprocess $\gamma g \rightarrow c {\bar c}$ . Hence,  SSA in 
$e + p^\uparrow \rightarrow e +J/\psi +X$, if observed, can be used as a
clean probe of gluon Sivers function. In addition, charmonium production mechanism  
can also have implications for this SSA and therefore, its study can help     
probe the production mechanism for charmonium.

\section{Estimate of the Sivers Asymmetry}

There are several models for charmonium production. We use the color
evaporation model (CEM)  as its simplicity makes it 
suitable for an initial study of SSA in the charmonium production.
This model was first proposed by Halzen and
Matsuda \cite{hal} and Fritsch \cite{fri}. In this model,  a statistical 
treatment of color is made
and the probability of finding a specific quarkonium state is assumed to be
independent of the color of heavy quark pair. In later versions of this model it has
been found that the data are better fitted if a phenomenological factor is included
in the differential cross section formula, which depends on a Gaussian distribution
of the transverse momentum of the charmonium \cite{ce2}.
We have used Weizsacker-Williams equivalent photon approximation for the
photon distribution of the electron \cite{wwf1,wwf2}, to
calculate the cross section for the process 
$e+\pup\rightarrow e+ J/\psi+X$ at low virtuality of the photon. The underlying  partonic process at
LO is $\gamma g\rightarrow c\bar{c}$ and therefore,
the only $\kt$ dependent pdf appearing is the gluon Sivers function.
For a complete calculation of photoproduction of $J/\psi$ one has to
consider higher order contributions and also the resolved photon  contributions
\cite{ce2}. Also the gauge links or Wilson lines present in the TMD
distributions are important at higher order \cite{feng}.

According to CEM, the cross section for charmonium production is
proportional to the
rate of production of $c\bar{c}$ pair integrated over the mass range $2m_c$
to $2m_D$ 
\be
\sigma=\frac{1}{9}\int_{2m_c}^{2m_D} dM \frac{d\sigma_{c\bar{c}}}  
{dM}
\ee  
where $m_c$ is the charm quark mass and $2m_D$ is the $D\bar{D}$ threshold,
$M^2$ is the squared invariant mass of the $c {\bar c}$ pair.

To calculate SSA in scattering of electrons off a polarized proton target,
we assume
a generalization of CEM expression by taking into account the transverse
momentum dependence
of the Weizsacker-Williams  (WW) function and gluon distribution function. The
numerator of the SSA can be written as
\bea
\frac{d^{4}\sigma^\uparrow}{dydM^2d^2\bfq_T}-\frac{d^4\sigma^\downarrow}
{dydM^2d^2\bfq_T}=
\frac{1}{s}\int [d^2\bfk_{\perp\gamma}d^2\bfk_{\perp g}]
\Delta^{N}f_{g/p^{\uparrow}}(x_{g},\bfk_{\perp g})
f_{\gamma/e}(x_{\gamma},\bfk_{\perp\gamma}) \nonumber\\
\times\>\delta^2(\bfk_{\perp\gamma}+\bfk_{\perp g}-\bfq_T)
\hat\sigma_{0}^{\gamma g\rightarrow c\bar{c}}(M^2)
\label{num-ssa}
\eea
where $y$ is the rapidity and $q_T$ in the transverse momentum of the
charmonium; $\Delta^{N}f_{g/p^{\uparrow}}(x_{g},\bfk_{\perp g})$   
is the gluon Sivers function, $ f_{\gamma/e}(x_{\gamma},\bfk_{\perp\gamma})$ 
is the photon distribution of the electron, given in the WW approximation.
The denominator would have a similar expression involving the unpolarized
gluon distribution of the proton; $f_{g/p}(x_g,\bfk_{\perp g})$, for which 
we use a gaussian form of $k_\perp$ distribution 
and a similar gaussian form for the transverse momentum dependence of the WW
function. To extract the asymmetry produced by the Sivers function, we
calculate the weighted asymmetry ~\cite{vogelsang-weight} 
\be
A_N^{\sin({\phi}_{q_T}-\phi_S)} =\frac{\int d\phi_{q_T}
[d\sigma ^\uparrow \, - \, d\sigma ^\downarrow]\sin({\phi}_{q_T}-\phi_S)}
{\int d{\phi}_{q_T}[d{\sigma}^{\uparrow} \, + \, d{\sigma}^{\downarrow}]}
\label{weight-ssa}
 \ee
where ${\phi}_{q_T}$ and  $\phi_S$ are the azimuthal angles of 
the $J/\psi$ and proton spin respectively.  For the gluon Sivers function we
have used a model in our analysis, which has been  used in the literature
to calculate SSA in semi-inclusive deep inelastic scattering (SIDIS)
\cite{Anselmino-PRD72} and DY process \cite{Anselmino2009} 
(see \cite{us} for details). The parameters are taken from  
\cite{2011-parmeterization}. Other parameters we use are 

\hspace{3 cm}$\langle{k_{\perp g}^2}\rangle=\langle{k_{\perp \gamma}^2}
\rangle=0.25\>GeV^2$.

Also it is to be
noted that in the model we consider for charmonium production, namely the
color evaporation model, the only relevant scale is $M^2$ which is the
invariant mass of the heavy quark pair. This is integrated between a narrow
region, from $4 m_c^2$ to $4 m_D^2$ irrespective of the center-of-mass
energy of the experiment. So the scale evolution of the TMDs is not expected
to affect the asymmetry too much.

\begin{figure}  
\includegraphics[width=0.49\linewidth,angle=0]{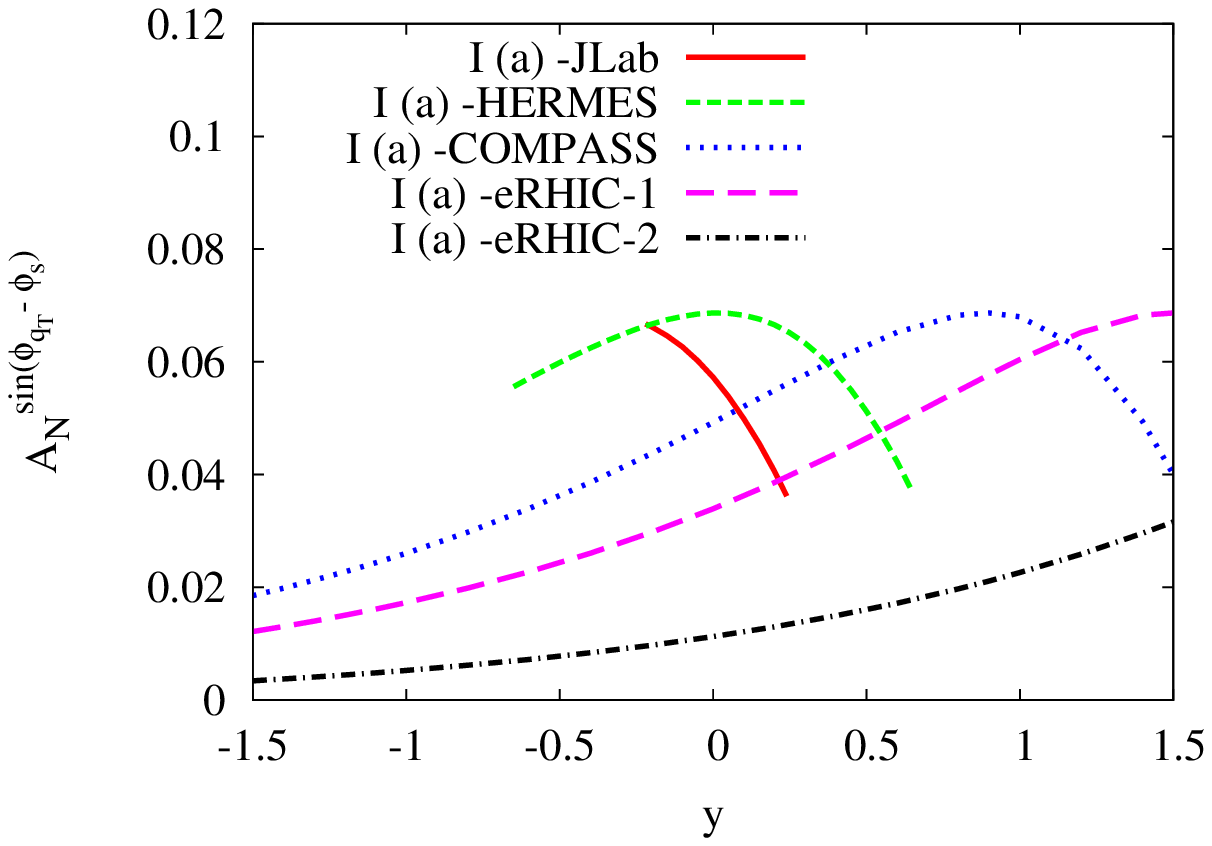}\hspace*{0.2cm}
\includegraphics[width=0.49\linewidth,angle=0]{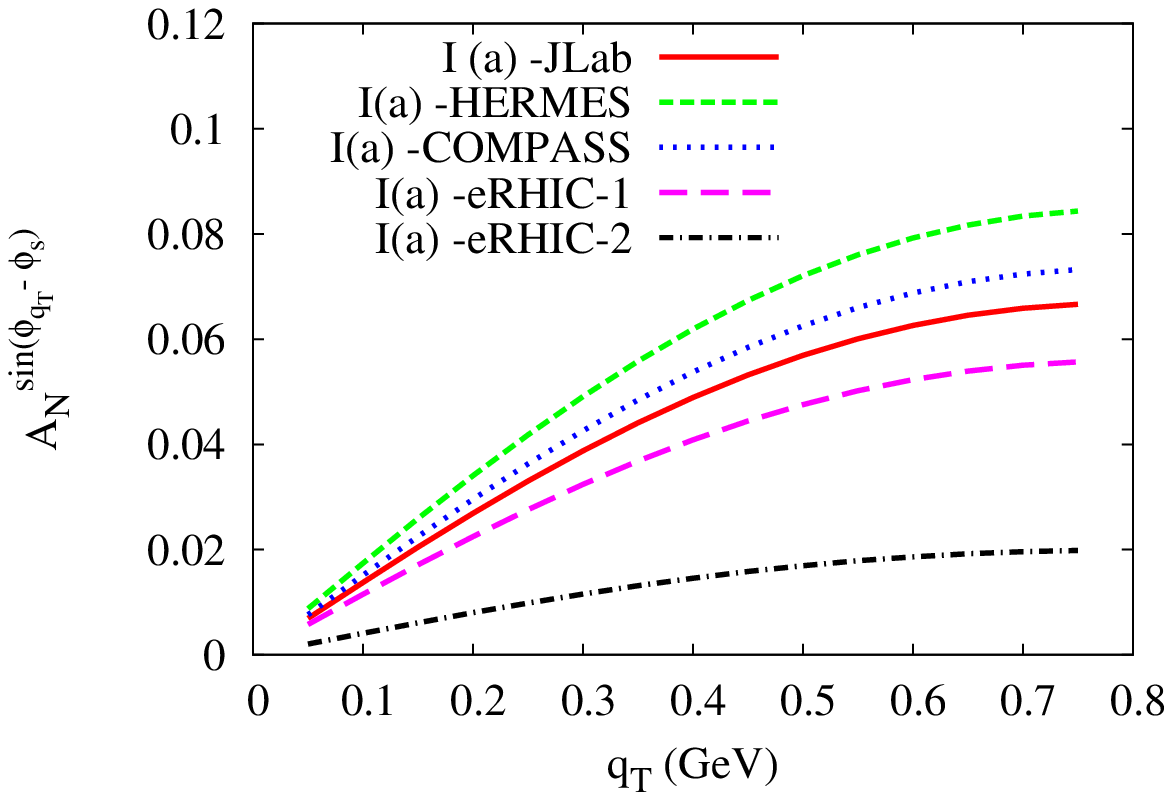}
 \caption{(Color online) The single spin asymmetry
$A_N^{\sin({\phi}_{q_T}-\phi_S)}$
for the $e + p^\uparrow \rightarrow e+J/\psi+X$ as a function of y (left
panel)  and $q_T$ (right panel). 
The plots are for model I (a) (see text) compared for JLab
 ($\sqrt s = 4.7$~GeV) (solid red line), 
HERMES ($\sqrt s = 7.2$~GeV) (dashed green line), COMPASS ($\sqrt s =
17.33$~GeV) (dotted blue line), 
eRHIC-1 ($\sqrt{s}=31.6$~GeV) (long dashed pink line) and eRHIC-2
($\sqrt{s}=158.1$~GeV) (dot-dashed black line).}
\end{figure}

In Fig. 1 we have shown  a comparison of the $y$ and $q_T$ dependence of the 
asymmetry at JLab, HERMES, COMPASS and eRHIC. Model  I refers to the
parametrization in \cite{Anselmino2009} and (a) refers to the
parametrization of the gluon Sivers function in terms of an average of the u
and d quark Sivers function \cite{us}.  Different experiments cover
different kinematical regions, and our results clearly show that the
asymmetry is sizable, and  that it is  worthwhile to look at SSA's in
charmonium production in order to extract information on the gluon Sivers
function.

\section{Acknowledgements}

A. Mukherjee thanks the organizers of XX th International Workshop of Deep
Inelastic Scattering and Related Areas.  
R.M.G. wishes to acknowledge support from the Department of Science and
Technology, India under Grant No. SR/S2/JCB-64/2007 under the J.C. Bose
Fellowship scheme. A. Misra and V.S.R. would like to thank Department of 
Science and Technology, India
for financial support under the grant No. SR/S2/HEP-17/2006 and to
Department of Atomic under the grant No. 2010/37P/47/BRNS.


\end{document}